\documentclass[twocolumn]{revtex4}

\input{epsf.tex}

\newcommand{\s}{\sigma}

\renewcommand{\a}{\alpha}

\newcommand{\be}{\begin{equation}}
\newcommand{\ee}{\end{equation}}
\newcommand{\bea}{\begin{eqnarray}}
\newcommand{\eea}{\end{eqnarray}}
\newcommand{\ba}{\begin{array}}
\newcommand{\ea}{\end{array}}
\def\J#1#2#3#4{{#1} {\bf #2}, #3 (#4)}
\def\PRD{Phys. Rev. D}

\def\PRL{Phys. Rev. Lett.}

\def\PTP{Prog. Theor. Phys.}

\def\APNY{Ann. Phys. (N.Y.)}

\def\JMP{J. Math. Phys.}
\def\CPAM{Comm. Pure Appl. Math.}

\def\CQG{Class. Quantum Grav.}

\def\PLA{Phys. Lett. A}

\def\IJMPD{Int. J. Mod. Phys. D}

\begin{document}
\draft
\title{Black hole-naked singularity dualism and the
repulsion\\ of two Kerr black holes due to spin-spin interaction}

\author{V.~S.~Manko\,$^\dag$ and E.~Ruiz$\,^\ddag$}
\address{$^\dag$Departamento de F\'\i sica, Centro de Investigaci\'on y de
Estudios Avanzados del IPN, A.P. 14-740, 07000 Ciudad de M\'exico,
Mexico\\$^\ddag$Instituto Universitario de F\'{i}sica Fundamental
y Matem\'aticas, Universidad de Salamanca, 37008 Salamanca, Spain}

\begin{abstract}
We report about the possibility for interacting Kerr sources to
exist in two different states -- black holes or naked
singularities -- both states characterized by the same masses and
angular momenta. Another surprising discovery reported by us is
that in spite of the absence of balance between two Kerr black
holes, the latter nevertheless can repel each other, which
provides a good opportunity for experimental detection of the
spin-spin repulsive force through the observation of astrophysical
black-hole binaries.
\end{abstract}


\maketitle


{\it Introduction.}---The recent direct detection of gravitational
waves by the LIGO and Virgo teams \cite{Abb} has aroused much
interest in the far-reaching study of binary black hole systems.
Though the numerical simulations are obviously the main tool for
modeling the generic mergers of two black holes, the analytical
results have also been obtained under some simplifying assumptions
\cite{EMa}. It appears that an important physical information
pertinent to the case of the head-on collisions of black holes was
also supplied by the well-known double-Kerr solution \cite{KNe} --
the two black holes cannot be in stationary equilibrium
\cite{Hoe,MRu}, the usual interpretation for which (see the
conjecture \cite{Wein} on the positiveness of the interaction
force) is that the spin-spin repulsion in a binary black hole
system is insufficient to counterbalance the gravitational
attraction. This could probably make one think that the end state
of a pair of colliding Kerr black holes must inevitably be a
single rotating black hole \cite{Ker} whose mass and angular
momentum would be equal to the aggregate masses and angular
momenta of the colliding constituents, as was surmised by Hawking
\cite{Haw} long ago. However, as will be shown in the present
letter, a different scenario of the dynamical evolution of a
binary black hole system is still possible and is based on the
recent results \cite{MRu2} obtained for the configurations of two
equal Kerr sources. We remind that in Ref.~\cite{MRu2} it has been
established that in a highly nonlinear regime the masses and
angular momenta may cease to define a binary configuration
uniquely, but the only example given in \cite{MRu2} treated the
case of hyperextreme sources, not black holes. Lately nonetheless
we have eventually succeeded in finding physically meaningful
examples of nonuniqueness for the black hole constituents too, and
the remarkable feature of the newly found binary black-hole
configurations is that the black holes in them can {\it repel}
each other due to prevailing of the spin-spin repulsion over
gravitational attraction, which prevents the two black holes from
merging into a single black hole. Moreover, we have discovered
that the nonuniqueness also includes a surprising phenomenon when
the same mass and angular momentum determine simultaneously a
black-hole and a hyperextreme configuration of two Kerr sources,
which allows to speak about the {\it black hole-naked singularity}
(BH/NS) {\it dualism} as a new highly nonlinear effect occurring
in the binary systems subject to a very strong gravity. We now
turn to the details.

{\it The configurations of two equal Kerr sources.}---The family
of two equal Kerr sources kept apart by a massless strut is
described by the metric \cite{MRu2}
\bea d s^2&=&f^{-1}[e^{2\gamma}(d\rho^2+d z^2)+\rho^2
d\varphi^2]-f(d t-\omega d\varphi)^2, \nonumber\\
f&=&\frac{A\bar A-B\bar B}{(A+B)(\bar A+\bar B)}, \quad
e^{2\gamma}=\frac{A\bar A-B\bar B} {16\lambda_0\bar\lambda_0
R_1R_2R_3R_4}, \nonumber\\ \omega&=&\omega_0-\frac{2{\rm
Im}[G(\bar A+\bar B)]}{A\bar A-B\bar B}, \nonumber\\
A&=&(R_1-R_2)(R_3-R_4)
-4\s^2(R_1-R_3)(R_2-R_4), \nonumber\\
B&=&2s\s[(1-2\s)(R_1-R_4)-(1+2\s)(R_2-R_3)], \nonumber\\
G&=&-zB
+s\s[2R_1R_3-2R_2R_4-4\s(R_1R_2-R_3R_4)\nonumber\\
&&-s(1-4\s^2)(R_1-R_2-R_3+R_4)], \label{mfn} \eea
where the functions $R_i$ are defined by the expressions
\bea R_i&=&X_i
\sqrt{\rho^2+(z-\a_i)^2}, \nonumber\\
X_1&=&-1/X_4=\phi(\mu+\sqrt{\mu^2-1}), \nonumber\\
X_2&=&-1/X_3=\phi(\mu-\sqrt{\mu^2-1}), \nonumber\\
\a_1&=&-\a_4=s\left(\frac{1}{2}+\s\right), \quad
\a_2=-\a_3=s\left(\frac{1}{2}-\s\right), \nonumber\\
\s&=&-\frac{i\sqrt{\mu^2-1}}{\nu}, \nonumber\\
\mu&=&\frac{(\phi^2+1)[\phi(\nu^2-4)+i\nu(\phi^2-1)]}
{2[(\phi^2+1)^2-i\nu\phi(\phi^2-1)]}, \label{Rrn} \eea
and the constants $\lambda_0$ and $\omega_0$ have the form
\bea \lambda_0&=&\frac{1}{\nu^2}
(1-\mu^2)[(1+\phi^2)^2-\nu^2\phi^2], \nonumber\\
\omega_0&=&\frac{2is(1-\phi^4-2i\mu\nu\phi^2)}
{(1+\phi^2)^2-\nu^2\phi^2}. \label{K0} \eea

The above formulas (\ref{mfn})-(\ref{K0}) involve three arbitrary
parameters, which are the coordinate distance $s$ between the
centers of the Kerr constituents (see Fig.~1), the real positive
constant $\nu$, and the unitary complex parameter $\phi$, so that
$\phi\bar\phi=1$; the black-hole configurations correspond to real
positive $\s$, while the pure imaginary $\s$ describe the binary
systems of hyperextreme Kerr objects (naked singularities).
Instead of $\nu$ and $\phi$, however, one may use the
dimensionless individual mass $m$ and angular momentum per unit
mass $a$ of each constituent (these are related to the total
dimensional quantities by the relations $M_T=2ms$, $J_T=2mas^2$),
and finding $\phi$ in terms of $m$ and $a$ then reduces to solving
the cubic equation
\bea &&Z^3-pZ^2+\bar pZ-1=0, \quad Z\equiv\phi^2, \nonumber\\
&&p\equiv p_x+ip_y=-\frac{m+2}{m}-\frac{i(2m+1)^2} {ma},
\label{eqZ} \eea
while the dependence $\nu(m,a)$ is determined by the equation
\be \nu=-\frac{i[m(\phi^2+1)^2+2\phi^2]} {m\phi(\phi^2-1)},
\label{nu} \ee
into which the explicit form of $\phi$ should be inserted after
the resolution of equation (\ref{eqZ}), taking into account that
the correct choice of sign in $\phi=\pm\sqrt{Z}$ must ensure
$\nu>0$. Obviously, for $m$ and $a$ the following relations hold:
\be m=-\frac{2}{p_x+1}, \quad a=\frac{(p_x-3)^2}{2p_y(p_x+1)}.
\label{map} \ee

As has been established in \cite{MRu2}, equation (\ref{eqZ}) may
have, for given values of $m$ and $a$, up to three unitary roots,
and the nonuniqueness zone is defined by the ``tanga curve''
depicted in Fig.~2, inside of which (and on the curve itself) the
values of $m$ and $a$ do not lead to unique configurations of the
Kerr constituents. The only example of nonuniqueness considered in
\cite{MRu2} involved the hyperextreme Kerr sources because at that
time we did not yet know how to extend our analysis to the
black-hole case. The difficulty that exhibits the latter case is
the need, in addition to finding configurations with real positive
$m$ and $\s$, to satisfy also the condition
$\s^2<\textstyle{\frac{1}{4}}$ ensuring the separation of the
black holes on the symmetry axis. The strategy that eventually
allowed us to get the desired two different binary black-hole
configurations defined by the same values of $m$ and $a$ consisted
in first identifying one physically meaningful black-hole
configuration on the tanga curve (where only two unitary $Z$ are
possible) and then searching the roots of the cubic equation
(\ref{eqZ}) inside the curved triangle which would be slightly
different from the boundary values. Below we give the numerical
value of $p$ found in this way which determines two different
black-hole configurations and one hyperextreme configuration of
two Kerr constituents, all three configurations sharing the same
mass and angular momentum:
\be p=-1.146560133416505 + 2.152383532475482 i. \label{pv} \ee
The corresponding values of $m$ and $a$ obtainable from
(\ref{map}) are the following:
\be m=13.64628, \quad a=-27.25276, \label{mav} \ee
and these are given up to five decimal places (like all the
remaining quantities hereafter). Then the first binary
configuration of black holes is described by the set of the
parameters
\bea \phi&=&-0.67856-0.73455i, \quad \nu=1.35343, \nonumber\\
\mu&=&-0.83296, \quad \s=0.40884, \nonumber\\
X_1&=&0.97166+0.23637i, \nonumber\\ X_2&=&0.15875+0.98732i,
\label{phiv1} \eea
while the second binary black-hole configuration is defined by the
values
\bea \phi&=&-0.67819-0.73489i, \quad \nu=1.35145, \nonumber\\
\mu&=&-0.86653, \quad \s=0.36933, \nonumber\\
X_1&=&0.954475+0.29829i, \nonumber\\ X_2&=&0.22086+0.97531i.
\label{phiv2} \eea
Remarkably, the above two subextreme configurations are also
accompanied by a binary system of hyperextreme Kerr sources,
namely,
\bea \phi&=&-0.07962-0.99683i, \quad \nu=0.08623, \nonumber\\
\mu&=&-2.07558, \quad \s=-21.09186i, \nonumber\\
X_1&=&0.02044+0.25597i, \nonumber\\ X_2&=&0.31007+3.88202i,
\label{phiv3} \eea
so that, on the one hand, we have demonstrated the nonuniqueness
of binary black-hole configurations at certain values of masses
and angular momenta, and on the other hand have established the
subextreme/hyperextreme duality of the binary configurations which
means that in a highly nonlinear regime a pair of interacting Kerr
sources can exist either as a black-hole binary or a system of two
naked singularities, both states characterized by the same masses
and angular momenta of the constituents.

From (\ref{mav}) it follows that $J_T^2/M_T^4=0.99709$, and hence
all the configurations (\ref{phiv1})-(\ref{phiv3}) would resemble
a nearly extreme single Kerr black hole to a distant observer.

{\it The interaction force.}---An outstanding feature of the
particular configurations presented above is that the spin-spin
repulsion in them exceeds the gravitational attraction, so that
the constituents repel each other. This can be seen by analyzing
the formula of the interaction force which has the form
\cite{Isr,Wei}
\be {\cal F}= \frac{1}{4}\left(e^{-\gamma_0}-1\right), \label{F}
\ee
with $\gamma_0$ denoting the value of the metric function $\gamma$
on the part of the symmetry axis separating the two constituents.
Then, taking into account that
\be e^{2\gamma_0}=g_+^2/g_-^2, \quad g_\pm=\phi^4
\pm(\nu^2-2)\phi^2+1, \label{g0} \ee
we get for the three configurations (\ref{phiv1})-(\ref{phiv3}),
labeled respectively with subindices I, II and III, the following
values of ${\cal F}$:
\be {\cal F}_{I}=-0.24235, \quad {\cal F}_{II}=-0.24, \quad {\cal
F}_{III}=-0.24887, \label{Fi} \ee
and one can see that these values differ only insignificantly from
each other. This means that the specific type of a Kerr source --
black hole or naked singularity -- does not really have any
serious effect on the interaction force in a binary configuration
if the subextreme and hyperextreme constituents have the same
masses and angular momenta and are separated by the same
coordinate distance.

{\it Physical implications.}---The existence of the repulsion
between two subextreme Kerr black holes gives a clear indication
that the spin-spin interaction should not be neglected in the
analysis of the head-on collision of corotating black holes
because this interaction is able to prevent the colliding
black-hole binary from forming a single black hole. Indeed, our
findings suggest that the following dynamical scenario for such a
collision should not be excluded in principle: The two initially
separated black holes approach each other under the action of the
gravitational attraction which is inversely proportional to $s^2$;
the contribution of the spin-spin force at large separation is
negligible, being as is well known \cite{Wal} inversely
proportional to $s^4$. However, at a small separation, the
spin-spin repulsion becomes dominant, and the black holes will
bounce off each other, moving away from each other to a distance
where the gravitational attraction prevails again and provokes a
next cycle of approaching and bouncing of the black holes. Since
such a nonstationary process is axially symmetric, no angular
momentum is radiated away, while the loss of mass through
gravitational radiation in the case of separated colliding black
holes is insignificant \cite{CSBC}, and hence we end up with a
binary system of oscillating black holes, that loses very slowly
its mass, as a possible final state of the head-on collision.

Apparently, the nonlinear effects are strongest at the smallest
separation distances between the components of the binary system,
when the BH/NS dualism is most likely to show itself up. Actually,
it can be shown by passing to the dimensional quantities that the
large $s$ approximation withdraws the binary system from the
nonuniqueness zone, which explains in particular why the dualism
phenomenon has not been discovered earlier. It is also clear that
at short distances the individual stationary limit surfaces of the
interacting Kerr constituents may form a combined ergoregion, as
it happens for instance in the configurations
(\ref{phiv1})-(\ref{phiv3}). Interestingly, the binaries with
attracting constituents (${\cal F}>0$) are then distinguishable
from the configurations with repelling constituents (${\cal F}<0$)
by the presence of a massless ring singularity off the symmetry
axis in the latter configurations which has a `benign character'
using Wald's terminology \cite{Wal2} and just signals that the
common ergosurface is about to be divided into two disconnected
parts or has already started such a division. We emphasize that in
the fully dynamical binary configurations, which are mimicked by
our solutions and could be simulated in the framework of numerical
relativity \cite{DOr}, neither of the two aforementioned
singularities would appear.

Let us also mention the relation of our analytical results to the
discussion held during the last decade in the framework of a test
particle approximation about the possibility to convert a spinning
black hole into a naked singularity
\cite{JSo,BCK,LBa,Duz,Wal3,Duz2}. The similarity between the paper
\cite{JSo} and our work is that the black holes in both approaches
must be almost extreme for being able to be turned into the naked
singularities, the transition itself involving a very fine tuning
of the parameters. At the same time, our research based on the
exact solution of Einstein's equations specifies and clarifies
various important aspects connecting the black holes and naked
singularities that cannot be taken into account by the
approximation scheme of Refs.~\cite{JSo,BCK,LBa,Duz,Wal3,Duz2}.
First of all, the binary systems in our analysis involve
interacting black holes, with full account of the spin-spin
repulsion contribution, when the values of the individual angular
momenta defining the extremality condition can exceed considerably
the angular momentum of a single Kerr extreme black hole
\cite{MMR} (the analogous change of the extremality condition for
binaries in electrostatics was discussed in \cite{Emp}). As a
consequence, for the configurations (\ref{phiv1})-(\ref{phiv3})
the ratio $|a|/m$ is equal to 1.99712, being almost twice greater
than that of a single extreme black hole, which illustrates well
that in the binary systems the inequality $|a|/m>1$ may determine
both the naked singularities and the rapidly rotating black holes.
Moreover, it is really remarkable that the same values of $m$ and
$a$ can define a pair of subextreme or hyperextreme constituents
characterized by the presence or absence of the event horizons,
and it seems that the transition from the black-hole to the
naked-singularity state (and vice versa) in such binaries, to
which in particular belong the configurations
(\ref{phiv1})-(\ref{phiv3}), presumably occurs in a spontaneous
way. In this respect we would like to remark that the conversion
of a Kerr naked singularity into a black hole by means of test
bodies would seem to us a simpler process than destroying the
black hole horizon through overspinning discussed in \cite{JSo},
since the former requires exclusively an increase in mass, and
hence it would be plausible to suppose that some astrophysical
black holes might have been initially formed as naked
singularities and only later became black holes after consuming a
necessary amount of surrounding matter. Let us also note for
completeness that in the extreme limit ($\s=0$) the metric
(\ref{mfn}) reduces to a special subfamily of the well-known
Kinnersley-Chitre solution \cite{KCh} determining two identical
corotating extreme black holes which was identified and analyzed
in the paper \cite{MRu3}. Apparently, the degenerated solution is
unsuitable for treating any nonextreme configurations, and in
particular, if used in a gedanken experiment to destroy the
horizons of extreme black holes with test bodies, it would give
the same negative result as earlier obtained by Wald \cite{Wal4}
with the aid of a single Kerr extreme solution. At the same time,
the extreme binary configurations from \cite{MRu3} can be shown to
provide analytical examples of two repelling extreme black holes,
thus supporting our results concerning the more general
non-extreme case.

Therefore, we have shown that the spin-spin interaction in binary
configurations of Kerr sources can play an important role at the
last stages of the merging or colliding processes, giving rise to
various interesting nonlinear effects. This interaction turns out
to be a major factor able to prevent two colliding corotating
black holes from forming a single black hole; alternatively, it
may lead to the formation of the two-component oscillators as the
end state of the collision, and in principle the latter binary
oscillating systems might be detectable through the astrophysical
observations. Our results also suggest that taking into account of
the spin-spin interaction could cause an elongation in time of the
gravitational wave signals which are being received from the
generic merging black holes or neutron stars. Lastly, the present
research clearly demonstrates that the physics of binary black
holes is richer than that of single black hole spacetimes, and we
expect that a clever amalgamation of the numerical and analytical
approaches to the study of the binary configurations will be able
to shed more light on the properties of the interacting black
holes in the future.


We are thankful to Emanuele Berti for useful comments. We are also
very grateful to the anonymous referee for pointing out a misprint
in the expression for $\lambda_0$ in the original version of the
paper. This work was partially supported by the CONACYT of Mexico,
and by Project FIS2015-65140-P (MINECO/FEDER) of Spain.

\begin{figure}[htb]
\epsfysize=60mm\epsffile{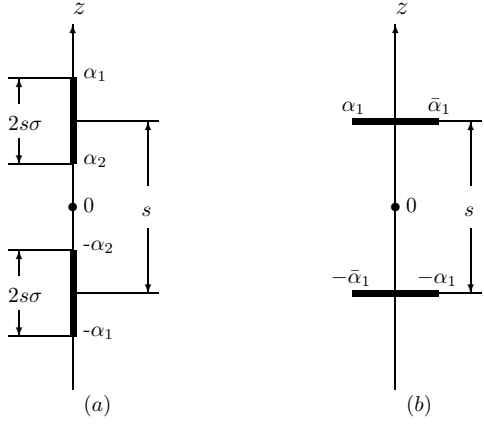} \caption{Two possible
configurations of a pair of equal nonextreme Kerr sources: ($a$) a
black-hole binary, ($b$) a system of two naked singularities.}
\end{figure}

\begin{figure}[htb]
\epsfysize=70mm\epsffile{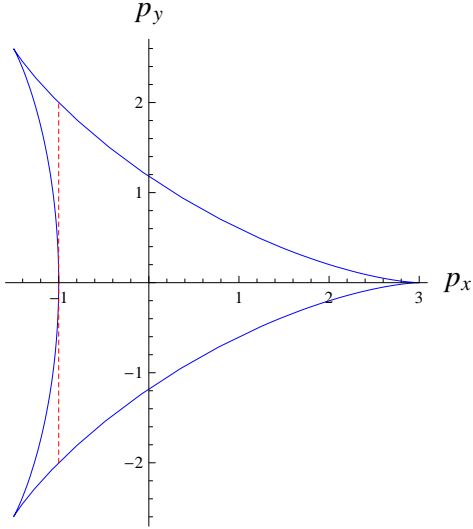} \caption{The curved triangle
separating the unique configurations with particular $m$ and $a$
from those for which the same particular values of mass and
angular momentum are shared by three different configurations (the
latter lie inside the tanga curve).}
\end{figure}

\end{document}